\documentclass[aps,preprint,prd,showpacs,nofootinbib]{revtex4}
\usepackage{amsmath}
\usepackage{graphicx}
\usepackage{subfigure}
\usepackage{bm}
\usepackage{amssymb}
\usepackage{mathtools}
\usepackage{enumerate}
\usepackage{color}
\usepackage[colorlinks,linkcolor=magenta,anchorcolor=cyan,citecolor=blue]{hyperref}

\newcommand{\bal}{\begin{aligned}}
\newcommand{\eal}{\end{aligned}}
\newcommand{\R}{\mathcal{R}}
\newcommand{\h}{\mathcal{H}}
\newcommand{\dn}{\delta N}
\newcommand{\e}{\eta}
\newcommand{\z}{\zeta}

\begin{document}

\title{Bounce in GR and higher-order derivative operators}

\author{Gen Ye$^{1}$\footnote{yegen14@mails.ucas.ac.cn}}
\author{Yun-Song Piao$^{1,2}$\footnote{yspiao@ucas.ac.cn}}

\affiliation{$^1$ School of Physics, University of Chinese Academy of
    Sciences, Beijing 100049, China}

\affiliation{$^2$ Institute of Theoretical Physics, Chinese
    Academy of Sciences, P.O. Box 2735, Beijing 100190, China}

\begin{abstract}

Recent progress seems to suggest that one must modify General
Relativity (GR) to stably violate the null energy condition and
avoid the cosmological singularity. However, with higher-order
derivative operators of the scalar field (a subclass of the degenerate
higher-order scalar-tensor theory), we show that at energies well
below the Planck scale, fully stable nonsingular cosmologies can
actually be implemented within GR.

\end{abstract}
\maketitle

\section{Introduction}

It is well-known that General Relativity (GR) suffers the
singularity problem, which indicates that our understanding about
gravity and the origin of the universe is incomplete
\cite{Hawking1970:singularity,Guth2003:inflationary}. It is
still an elusive task to look for an ultraviolet
(UV)-complete theory to describe what happens at the
"singularity". However, searching for fully stable nonsingular
cosmologies with the effective field theory (EFT), which captures low
energy behaviors of the complete theory, might be an alternative approach.

In spatially flat nonsingular cosmologies, the Null Energy
Condition (NEC) must be violated for a period. However, it is
often accompanied by (ghost, gradient) instabilities
\cite{Rubakov2016:generalized,Kobayashi2016:generic}, or singularities (strong coupling) in the perturbed action, see
also
Refs.\cite{Easson:2011zy,Ijjas:2016tpn,Ijjas2017:fully,Dobre2018:unbraiding}. Recently, it has been found that fully stable
nonsingular cosmological solutions do exist in the EFT beyond Horndeski
\cite{Cai:2016thi,Creminelli:2016zwa,Cai:2017tku,Cai:2017dyi,Kolevatov:2017voe,Mironov:2018oec}.
%Nevertheless, in the models built in the subclass of
Degenerate higher-order scalar-tensor (DHOST) theory
\cite{Langlois:2018dxi} actually is a rich pool for such EFTs
\cite{Ye:2019frg}. However, it is noteworthy that in the
nonsingular models built, the gravity has been no longer GR-like\footnote{By "GR", we refer to a theory where matter is minimally coupled to the Einstein-Hilbert action, i.e:
\[S=\int d^4x\sqrt{-g}\left(\frac{M^2_p}{2}R+L_m[g_{\mu\nu}]\right).\]
In particular, the scalar field is minimally coupled to the gravitational metric $g_{\mu\nu}$.}.

%the derivative coupling of the scalar field to the gravity must be

Recently, the LIGO Scientific and Virgo Collaborations have
detected the gravitational wave (GW) signals of binary black holes
(BH) \cite{Abbott:2016blz} and binary neutron star mergers
\cite{TheLIGOScientific:2017qsa}, which opened a new window to
probe the gravity physics. The results of all tests performed in
Refs.\cite{TheLIGOScientific:2016src,Abbott:2018lct} showed
perfect agreement with GR, particularly in the strong-field
regime.
%As a result, a large class of the extensions of GR has
%been highly disfavored,
%e.g.\cite{Creminelli2017:dark,Ezquiaga2017:dark,Baker:2017hug,Sakstein:2017xjx}.
%Currently, GR is the well-tested theory of gravity.
Currently, GR is still a well-established effective theory in the low
energy regime of the UV-complete theory, though it must break down
around the Planck energy.

How to implement the nonsingular bounce with GR? It is well-known
that the $P(\phi,X)$ theory can hardly bring a stable NEC
violation. To stably realize such a violation, one
may include higher-order derivative operators $(\Box\phi)^2$,
$(\phi^{\mu\nu})^2\cdots$ in the $P(\phi,X)$ theory, and set the
EFT as, e.g.\cite{ArkaniHamed:2003uy},
\begin{equation}\label{effective}
L\sim {M_P^2\over 2}\,^{(4)}R+P(\phi,X)+{\cal
O}\left((\Box\phi)^2, (\phi^{\mu\nu})^2\cdots\right).
\end{equation}
%which naturally capture the effect of high energy physics without
%needing not consider a particular UV-completion of
%\eqref{effective} \cite{deRham:2017aoj}.
%Such additional higher derivative operators were previously used
%in an attempt to regulate the sound speed in the simplest scalar
%field P(X) theories,
Generally, higher-order corrections ${\cal
O}\left((\Box\phi)^2, (\phi^{\mu\nu})^2\cdots\right)$ are generated if one integrates out the massive particles beyond the cutoff scale \cite{deRham:2017aoj,deRham:2018red}. One frequently studied case is ${\cal O}\sim (\Box\phi)^2$, see
e.g.\cite{Creminelli:2006xe,Li:2005fm,Buchbinder:2007ad}. However,
the corresponding EFT must beg unknown physics in the sufficiently
far past, otherwise the higher-order derivative operator will show
itself the Ostrogradski ghost. It's possible to include such
higher-order derivative operators in the so-called
DHOST theory \cite{Langlois:2015cwa,Langlois:2017mxy}, see also
\cite{Gleyzes:2014dya}, without introducing any Ostrogradski instability \cite{Motohashi:2016ftl,Motohashi:2014opa}.

Nevertheless, which operator in ${\cal O}\left((\Box\phi)^2,
(\phi^{\mu\nu})^2\cdots\right)$ is indispensable for achieving a
pathology-free bounce in GR is still not clear so far. In this
paper, we will propose a consistent \eqref{effective}-like EFT for spatially-flat fully stable nonsingular cosmologies. We, with
it, will discuss how to evade the No-go Theorem
\cite{Rubakov2016:generalized,Kobayashi2016:generic} plaguing the
cosmologists, and show a concrete example for the cosmological
bounce.

\section{DHOST theory with $c_T=1$}\label{sec:lagrangian}
\subsection{Reducing to GR}

We begin with the DHOST theory with $c_T=1$ ($c_T$ is the speed of
GWs) \cite{Langlois:2017dyl}
\begin{equation}\label{ct1 dhost cov}
\bal L^{DHOST}_{c_T=1}=&  P+ Q  \Box\phi + A^{(4)}R
+{1\over A}\left({6 A_X^2} -\left(A-XA_X\right)B-{X^2 B^2\over 8} \right) \phi^\mu \phi_{\mu \nu} \phi_\lambda \phi^{\lambda \nu} \\
&+ B\phi^\mu \phi^\nu \phi_{\mu \nu} \Box \phi+ \frac{B}{
A}\left(2A_X+{X B\over 2}\right)(\phi_\mu \phi^{\mu \nu }
\phi_\nu)^2, \eal
\end{equation}
where $\phi_\mu\equiv \nabla_\mu\phi$,
$\phi_{\mu\nu}\equiv\nabla_\nu\nabla_\mu\phi$ and
$X\equiv\phi_\mu\phi^\mu$. The coefficients $A$, $Q$ and $B$ only depend on $\phi$ and $X$. According to the classification in Ref.\cite{Achour:2016rkg}, theory \eqref{ct1 dhost cov} belongs to class Ia DHOST theories.
%$^{(4)}R$ is the four-dimensional Ricci scalar.
Generally, $B$ and $A$ are independent functions. However, if
$B=-\frac{4}{X}A_X$, $L^{DHOST}_{c_T=1}$ will reduce to the
$c_T=1$ beyond-Horndeski theory $L^{bH}_{c_T=1}$
\cite{Creminelli2017:dark}.

It is significant to notice that if setting $A=const.$ and $Q=0$,
$L^{bH}_{c_T=1}$ will reduce to GR, while
$L^{DHSOT}_{c_T=1}$ will become GR plus extra DHOST operators
(higher-order derivative operators). The latter is not covered by the beyond-Horndeski Lagrangian \cite{Gleyzes:2014dya} but belongs to a subclass of the $c_T=1$ DHOST
theory. Degenerate conditions required by the DHOST theory guarantee
that such a combination of higher-order derivative operators is
free of the Ostrodradsky ghost.
%As pointed out in
%Refs.\cite{Cai:2016thi}\cite{Creminelli:2016zwa}, the fully stable
%nonsingular cosmologies do exist if we go beyond Horndeski.
A \eqref{effective}-like EFT will be Ostrodradsky ghost-free,
only if the degenerate conditions are satisfied.

\subsection{Perturbation in DHOST theories with $c_T=1$}

We adopt the ADM metric,
\begin{equation}\label{adm metric}
ds^2=-N^2dt^2+h_{ij}(dx^i+N^idt)(dx^j+N^jdt),
\end{equation}
where $N$ is the lapse, $N^i$ is the shift and $h_{ij}$ is the
spatial metric. In the following we will work in the unitary gauge and use $\e=\phi$ as the time coordinate (assuming $\phi_{\mu}$ is timelike). In particular, in this gauge $\phi_\mu=\delta_\mu^0$ and the dynamics
of $\phi$ is absorbed into $N(\e)$, as
$\dot{\phi}\equiv{d\phi}/{d\e}=1$ (for any operator $\mathcal{O}$, $\dot{\mathcal{O}}$ refers to derivatives with respect to the clock
time $\phi$, or equivalently $\eta$, and not as usual to the cosmic time $t$).

Defining
\begin{equation}\label{BB}
B=-\frac{4}{X}A_X+A{\tilde B}, \end{equation} we have
\[  L^{DHOST}_{c_T=1}=L^{bH}_{c_T=1}+\Delta L,\]
where
\[
\bal \Delta L=&A{\tilde B}\phi^\mu \phi^\nu \phi_{\mu \nu} \Box
\phi+\left(-A{\tilde B}+2XA_X{\tilde B}-{X^2A{\tilde B}^2\over
8}\right)\phi^\mu \phi_{\mu \nu} \phi_\lambda \phi^{\lambda
\nu}\\&+\frac{X{\tilde B}}{2}\left(-\frac{4}{X}A_X+A{\tilde
B}\right)(\phi_\mu \phi^{\mu \nu } \phi_\nu)^2. \eal
\]
%we can make the metric dynamical and write the covariant
%scalar-tensor theory \eqref{ct1 dhost cov} into a form consisting
%of 3-geometric objects (i.e: extrinsic curvature tensor $K_{ij}$
%and intrinsic curvature tensor $\R_{ij}$) on the constant $\e$
%hypersurfaces.
In the unitary gauge, one has
$L^{bH}_{c_T=1}={\tilde P}+{\tilde
Q}K+A(\mathcal{R}-\mathcal{K}_2)$ \cite{Gleyzes:2014dya}, where
$\R\equiv h^{ij}\R_{ij}$ is the Ricci scalar on the spacelike
hypersurface, $K\equiv h^{ij}K_{ij}$ is the extrinsic curvature
and $\mathcal{K}_2\equiv K^2-K_{ij}K^{ij}$.
%with time coordinate $\e$ one has
%\[\phi_{\mu\nu}=-\Gamma^0_{\mu\nu}\] Thus, to cast $\Delta L$ into
%a 3-geometric form, one only needs the connection in metric
%\eqref{adm metric}
%\[\bal\phi_{00}&=-\Gamma^0_{00}=-\frac{1}{N}\left(N^iN^jK_{ij}+\dot{N}+N^i\partial_iN\right)\\\phi_{0i}=\phi_{i0}&=-\Gamma^0_{0i}=-\frac{1}{N}\left(N^jK_{ij}+\partial_iN\right)\\\phi_{ij}&=-\Gamma^0_{ij}=-\frac{1}{N}K_{ij}\eal\]
%where upper dots denotes time derivatives with respect to $\e$.
The DHOST operators follow
\begin{equation}\label{dhost operator}
\bal
\phi^\mu \phi^\nu \phi_{\mu \nu}&=-\frac{1}{N^5}\left(\dot{N}-N^i\partial_iN\right)\equiv-\frac{1}{N^5}N',\\
\Box \phi&=\frac{1}{N^3}\left(\dot{N}-N^i\partial_iN\right)-\frac{1}{N}K\equiv\frac{1}{N^3}N'-\frac{1}{N}K,\\
\phi^\mu \phi_{\mu \nu} \phi_\lambda \phi^{\lambda
\nu}&=-\frac{1}{N^8}\left(\dot{N}-N^i\partial_iN\right)^2+\frac{1}{N^6}(\partial
N)^2\equiv-\frac{1}{N^8}N'^2+\frac{1}{N^6}(\partial N)^2. \eal
\end{equation}
Thus we have
\[
\Delta L=-\frac{3A{\tilde B}^2}{8N^{12}}N'^2+\frac{A{\tilde
B}}{N^6}N'K-\frac{\tilde
B}{N^5}\left(\frac{A}{N}+NA_N+\frac{A{\tilde
B}}{8N^5}\right)(\partial N)^2
\]
where the equality $X=-1/N^2$ is used. Replacing $-{\tilde
B}/(2N^5)$ with ${\tilde B}$, we get the ADM form of
$L^{DHOST}_{c_T=1}$ \eqref{ct1 dhost cov}
\begin{equation}\label{ct1 dhost adm}
\bal L^{DHOST}_{c_T=1}=&{\tilde P}+{\tilde
Q}K+A(\R-\mathcal{K}_2)-\frac{3A{\tilde
B}^2}{2N^2}N'^2-\frac{2A{\tilde B}}{N}N'K\\&+{\tilde
B}\left(2\frac{A}{N}+2A_N-\frac{A{\tilde B}}{2}\right)(\partial
N)^2.  \eal
\end{equation}

%In the FRW metric, one has $h_{ij}\to a^2h_{ij}$, it amounts to
%make the substitution $(\partial N)^2\to(\partial N)^2/a^2$ in
%Lagrangian \eqref{ct1 dhost adm}.
%The Euler-Lagrangian equations for \eqref{ct1 dhost adm} are
%showed in Appendix-\ref{apdx:EoM}.

%These equations are obviously degenerate.

%\footnote{To simplify $\Delta L$ we have made a coefficient
%redefinition $\tilde{B}=-B/(2N^5)$ and then drop the tildes in the
%final result \eqref{ct1 dhost adm}.}

%\section{Scalar perturbation}

We will work with \eqref{ct1 dhost adm}. To study the stability of
perturbations, we expand $L^{DHOST}_{c_T=1}$ in \eqref{ct1
dhost adm} to second order. Defining the metric perturbation
\[N^i=\delta^{ij}\partial_j\psi,\qquad h_{ij}=a^2(\e)e^{2\z}\delta_{ij},\]
we have $\mathcal{L}^{(2)}=3a^3N\zeta\delta L+a^3\delta N\delta
L+a^3N\delta_2 L$ at quadratic order, where $\delta_2L$ refers to
the expansion of $L$ at second order. To proceed, we first expand
$K_i^{\ j}$ and $\R$, \begin{equation} \label{Kij} K_i^{\
j}=\frac{1}{N}\left[\left(\h+\dot{\z}-\h\frac{\dn}{N}\right)\delta_i^j-\frac{\delta^{jk}}{a^2}\partial_i\partial_k\psi\right]+\mathcal{O}(\delta
N^2),
\end{equation}
\begin{equation}\label{R3}
\R=-\frac{2}{a^2}\big[2\partial^2\z+(\partial\z)^2-4\z\partial^2\z\big]+\mathcal{O}(\z^3),
\end{equation} where $\h\equiv {da\over ad\e}=NH$, and $H$ is
the Hubble parameter. The kinetic term
in $L^{DHOST}_{c_T=1}$ \eqref{ct1 dhost adm} is contributed by
$-A\mathcal{K}_2-\frac{3A\tilde{B}^2}{2N^2}N'^2-\frac{2A\tilde{B}}{N}N'K$.
Considering (\ref{Kij}) and (\ref{R3}), one finds that
\[\mathcal{L}^{(2)}_{kinetic}=a^3\frac{A}{N}\left(-6\dot{\z}^2-6\tilde{B}\dot{\z}\delta\dot{N}-\frac{3\tilde{B}^2}{2}\delta\dot{N}^2\right)=-\frac{6a^3A}{N}(\dot{\z}+\tilde{B}\delta\dot{N}/2)^2 \]
is diagonal for $\tilde{\z}=\z+\tilde{B}\dn/2$. The coefficients of the operators $N'^2$ and
$N'K$ should satisfy a relation in the DHOST theory ($\beta_2=-6\beta_1^2$, see
e.g.Ref.\cite{Langlois:2017mxy}). As a result,
$\mathcal{L}^{(2)}_{kinetic}$ is necessarily diagonal. Confronting
${\tilde\zeta}$ with the constraint $\delta
L/\delta(\partial^2\psi)=0$, we get
\begin{equation}\label{quad L}
\mathcal{L}^{(2)}=a^3NA\left[U\dot{\tilde{\z}}^2-V\frac{(\partial\tilde{\z})^2}{a^2}\right]
\end{equation}
with
\begin{equation}\label{u}
U=\frac{\Sigma}{\gamma^2}+\frac{6}{N^2},
\end{equation}
\begin{equation}\label{v}
V=\frac{2}{aA }\frac{d}{d\e}\left(a\mathcal{M}\right)- 2,
\end{equation}
where
\begin{equation}\label{gamma}
\gamma\equiv \left({1\over N}+N\alpha_B\right)\h+\dot{\tilde B}/2,
\end{equation}
\begin{equation}\label{sigma1}
\Sigma\equiv\h^2\left[\alpha_K+6\left(\alpha_B^2-\frac{\gamma^2}{\h^2N^2}\right)+{9\alpha_B{\tilde
B}\over N}+\frac{3d(\alpha_B\h A{\tilde B})/d\e}{\h^2NA}\right],
\end{equation}
\begin{equation}\label{m}
\mathcal{M}\equiv\frac{1}{\gamma}\left[\big({A\over
N}+A_N\big)-A{\tilde B}/2\right].
\end{equation}
Following the notation in \cite{Langlois:2017mxy}, one sets $\alpha_B$ and $\alpha_K$ as the coefficients of
the operators $\delta K\dn$ and $\dn^2$ respectively,
%in particular, referring
%to Lagrangian \eqref{ct1 dhost adm}, we see that
\begin{equation}
\alpha_B=\frac{1}{4NA\h}(NL_{NK}+2\h
L_{NS}),\qquad\alpha_K=\frac{1}{NA\h^2}\left(L_N+\frac{N}{2}L_{NN}\right),
\end{equation}
where $\mathcal{S}\equiv K_{ij}K^{ij}$. It can be checked\footnote{In Ref.\cite{Langlois:2017mxy}, a different time parametrization is chosen such that $\bar{N}=1$.} that our calculation conform with Ref.\cite{Langlois:2017mxy} by setting $N=1$.

\section{Bounce in GR}
\subsection{Expelling No-go with higher-order derivative operators}\label{sec:no-go}

%One of the difficulties non-singular cosmology faced with is the
%no-go theorem presented in reference

%The No-go Theorem showed in
%Refs.\cite{Rubakov2016:generalized,Kobayashi2016:generic} states
%that the nonsingular cosmologies based on the Horndeski theory
%will inevitably be plagued with the instabilities.

In the Horndeski theory, fully stable nonsingular cosmological
solutions are prohibited, the so-called No-go Theorem
\cite{Rubakov2016:generalized,Kobayashi2016:generic}, see also
\cite{Kolevatov:2016ppi,Akama:2017jsa,Ijjas2018:space-time,Banerjee:2018svi}
for relevant studies. One way out is going beyond Horndeski,
as pointed out in Refs.\cite{Cai:2016thi,Creminelli:2016zwa}. In
particular, in the beyond-Horndeski subclass of the DHOST theory,
solutions of fully stable nonsingular cosmologies have been
found
\cite{Cai:2017dyi,Kolevatov:2017voe,Mironov:2018oec,Ye:2019frg}.

%Bouncing cosmologies encounters one more problem:
%$\gamma$-crossing, that $\gamma$ crosses zero at some finite time.
%Note that the quantity (denoted $\tilde{\gamma}$ here) used in the
%$\gamma$-crossing literature
%\cite{Ijjas2016:classically,Ijjas2017:fully,Ijjas2018:space-time}
%is related to \eqref{gamma} by $\tilde{\gamma}=2A\gamma$. It's now
%clear that both healthy $\gamma$-crossing and avoiding no-go call
%for non-zero $A_N$ and/or $B$.

Setting $A= M_P^2/2=const.$ in \eqref{ct1 dhost adm}, we have
\begin{equation}\label{ct1 dhost adm1} \bal
L^{DHOST}_{c_T=1,A=M_P^2/2}=&{\tilde P}+{\tilde Q}K+{M_P^2\over
2}(\R-\mathcal{K}_2)-\frac{3M_P^2{\tilde
B}^2}{4N^2}N'^2-\frac{M_P^2{\tilde B}}{N}N'K\\&+{\tilde
B}\left(\frac{M_P^2}{N}-\frac{M_P^2{\tilde B}}{4}\right)(\partial
N)^2,  \eal
\end{equation}
which also belongs to a subclass of the DHOST theory. Recall the
redefinition \eqref{BB} and replacement $-{\tilde B}/(2N^5)\to{\tilde B}$ in
Sect.\ref{sec:lagrangian}, then the coefficient $B$ in the covariant theory \eqref{ct1 dhost cov}
is related to the $\tilde{B}$ in \eqref{ct1 dhost adm1}
by $B=-M_P^2N^5 {\tilde B}$. It is also noticed that if
$A=const.$, ${\tilde Q}=0$ in \eqref{ct1 dhost adm1} is equivalent to $Q=0$ and
$P={\tilde P}$ in $L^{DHOST}_{c_T=1}$ \eqref{ct1 dhost cov}.
Thus if ${\tilde Q}=0$, \eqref{ct1 dhost adm1} is actually a
\eqref{effective}-like EFT.

The essence of the No-go proof is rewriting $V>0$ ($c_S^2>0$) in
\eqref{v} as the integral inequality, see \cite{Kobayashi:2019hrl}
for a review,
\begin{equation}\label{no-go int}
a\mathcal{M}\big|_f-a\mathcal{M}\big|_i>\int_{i}^{f}aAd\e.
\end{equation}
In the nonsingular models, the integral $\int_{i}^{f}aAd\e$ will
diverge, thus $\mathcal{M}$ must cross 0 at a certain time.
According to \eqref{m}, we have \begin{equation} \label{M1}
\mathcal{M}={M_P^2\over 2\gamma}\left({1\over N}-{\tilde
B}/2\right)
\end{equation} for \eqref{ct1 dhost adm1}. Thus we might get $\mathcal{M}=0$ by adjusting ${\tilde B}(N,\eta)$, or
equivalently $B(X,\phi)$ in $L^{DHOST}_{c_T=1}$ \eqref{ct1
dhost cov}. This suggests that it is possible to build fully
stable nonsingular cosmological models with \eqref{ct1 dhost
adm1} (equivalently, \eqref{effective}-like EFTs).

\subsection{An example}\label{sec:model}

To show that the observation made in Sect.\ref{sec:no-go} is
correct, we will present a concrete model for the nonsingular
bounce, which might have significant applications in early
universe scenarios,
e.g.\cite{Khoury2001:ekpyrotic,Piao:2003zm,Piao:2004me,Qiu:2011cy}.

We adopt
\begin{equation}\label{bkgrd}
\h=H/N=\frac{\e}{3(1+\e^2)},
\end{equation}
with $N(\e)=1$ as the background solution. When $\e<0$, the universe contracted with ${\cal H}\sim
1/\e<0$. Cosmological bounce happened at $\e=0$. We might set
${\tilde P}(N,\e)$ and ${\tilde B}(N,\e)$ in
$L^{DHOST}_{c_T=1,A=const.}$ \eqref{ct1 dhost adm1} as
\begin{equation}\label{free functions}
{\tilde P}(N,\e)=
\frac{g_1(\e)}{2N^2}+\frac{g_2(\e)}{N^4}+g_3(\e),
\end{equation}
and ${\tilde B}(N,\e)=g_4(\e)$. Here, since ${\tilde
Q}=0$, ${\tilde P}(N,\e)$ is actually equivalent to
$P(X,\phi)=g_1(\phi)X/2+g_2(\phi)X^2+g_3(\phi)$ in
$L^{DHOST}_{c_T=1}$ \eqref{ct1 dhost cov}.

One simple possibility for \eqref{sigma1} is, see also
\cite{Ye:2019frg},
\begin{equation}\label{nsgl u} \Sigma=c_1(\e)\gamma^2.
\end{equation}
According to Eq.\eqref{u}, we will have $U>0$ for a suitable
$c_1(\e)$. Combining Eq.\eqref{nsgl u} with the background
equations \eqref{eom1} and \eqref{eom2} in Appendix
\ref{apdx:EoM}, we get the algebraical solutions of $g_1(\e)$,
$g_2(\e)$ and $g_3(\e)$, see Appendix \ref{apdx:g123}.

Inserting ${\tilde B}(N,\e)=g_4(\e)$ into Eq.\eqref{gamma}, we
have $\gamma=\h+{\dot{g}_4}/{2}$. Thus \begin{equation}\label{m
model} \mathcal{M}=\frac{M^2_P(1-g_4/2)}{\dot{g}_4+2\h}.
\end{equation}
Requiring that around $\e=0$, $1-g_4/2=0$ and
$\dot{g}_4\sim {\cal H}$ (so ${\cal M}=0$), we consider such a $g_4$,
\begin{equation}\label{g4}
g_4(\e)=\int^{+\infty}_\e2\mu {\cal H}(s) e^{-s^2/\lambda^2}ds,
\end{equation}
with $\lambda$ set by
$g_4(0)={\mu}e^{1/\lambda^2}\Gamma(0,1/\lambda^2)/3=2$. Fig.\ref{H
g4} plots the evolutions of $\dot{g}_4$ for $\mu=0.9$ and $\cal
H$. When $|\e|\gg \lambda$, $g_4= 0$, we will have a $P(X,\phi)$
EFT with GR. Inserting \eqref{m model} into \eqref{v}, we have
$V(\e=0)=\frac{2(2\mu-1)}{-\mu+1}$, so $c_S^2(\e=0)=V/U>0$
suggests $0.5<\mu<1$.

\begin{figure}
    \centering
    \includegraphics[width=4in]{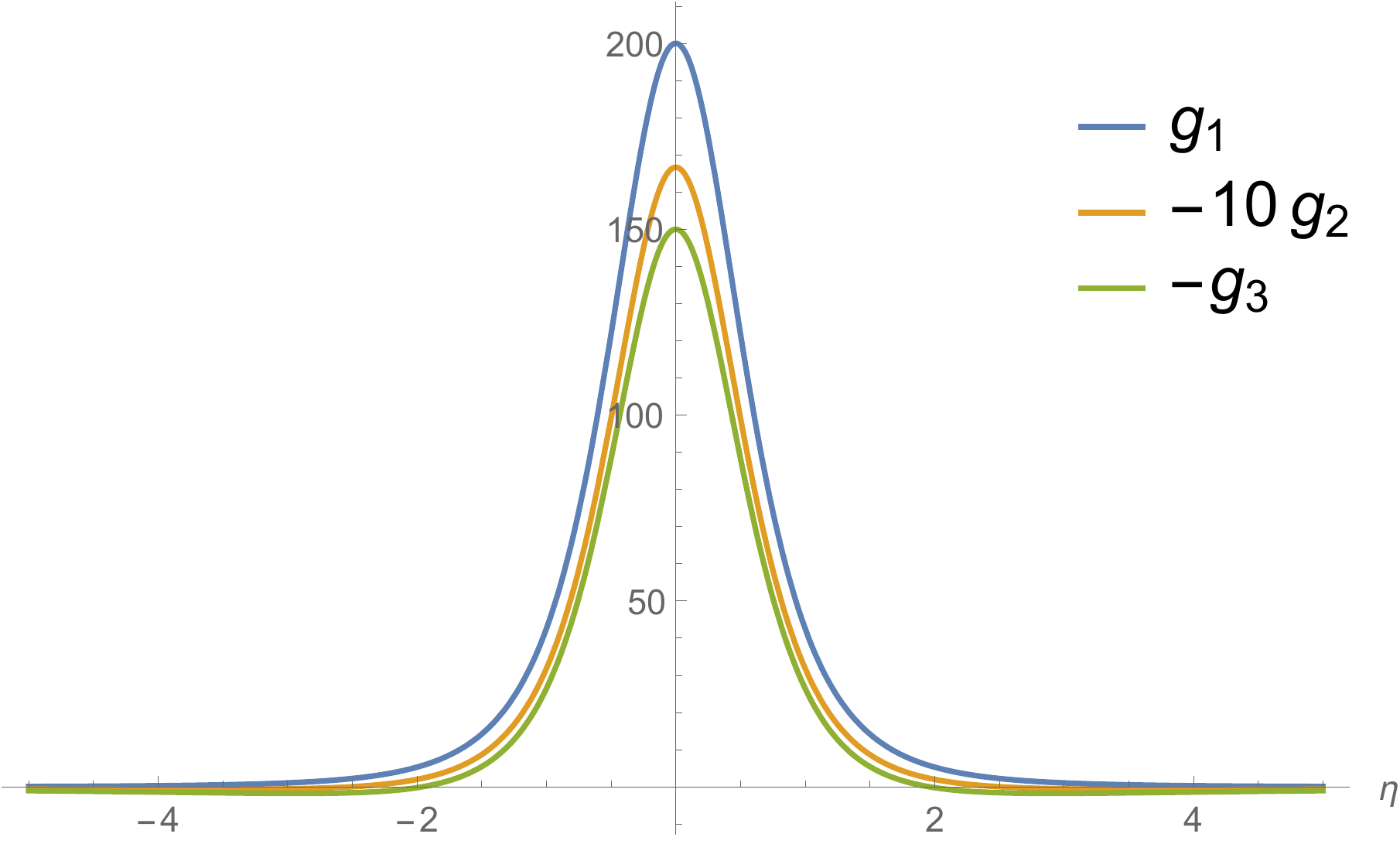}
    \caption{Coefficients $g_1$, $g_2$ and $g_3$ in ${\tilde P}(N,\e)$ \eqref{free functions}.}
    \label{g123}
\end{figure}

\begin{figure}
    \centering
   \includegraphics[width=4in]{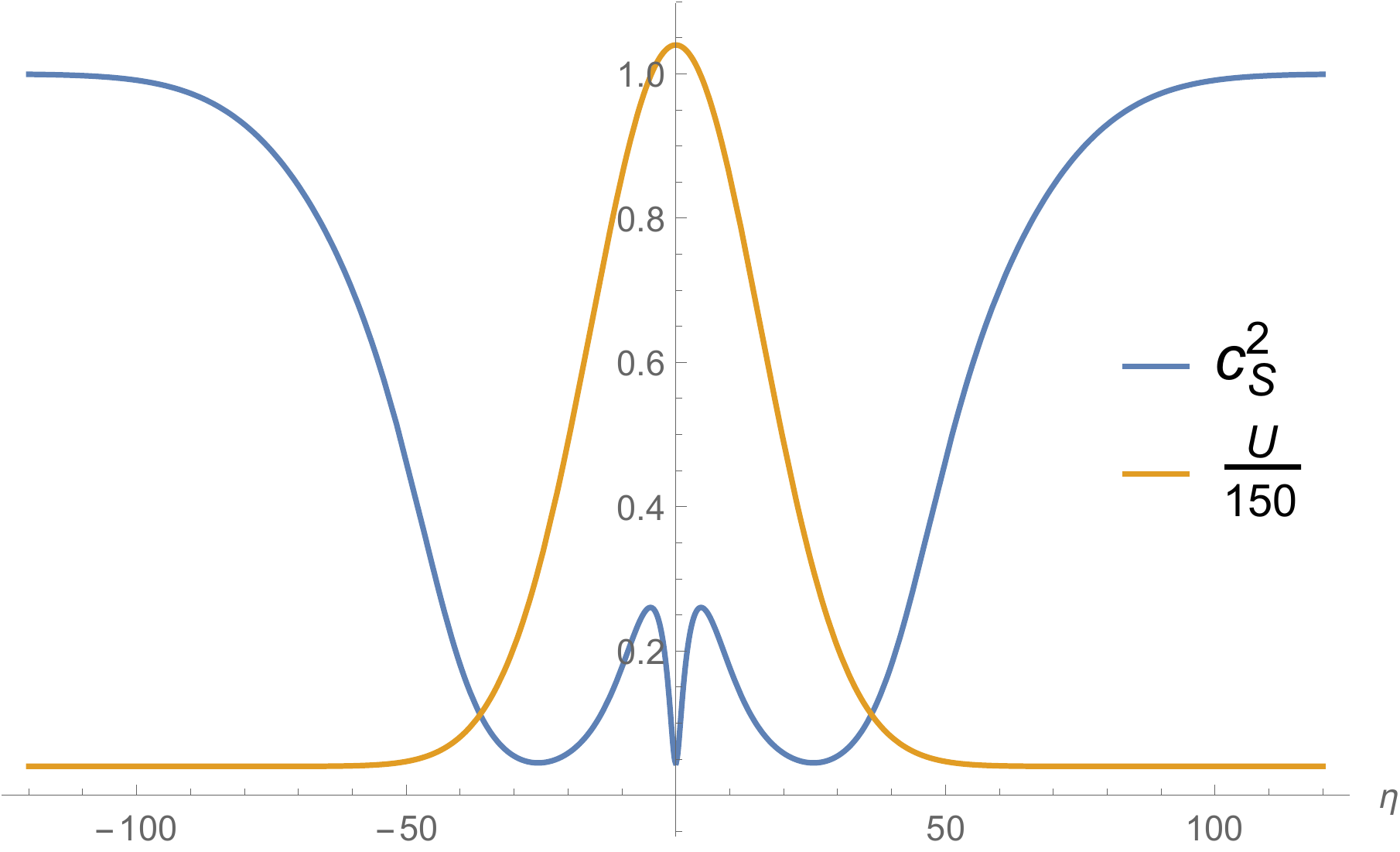}
    \caption{Throughout the whole evolution, $c_S^2>0$, while $c_S^2\to1$ as $|\e|\to\infty$.  }
    \label{ucs2}
\end{figure}

As a concrete example, we plot Figs.\ref{g123} and \ref{ucs2} with
$c_1(\e)=150e^{-\e^2/500}$. We see that the model is fully stable. As pointed out in Ref.\cite{Achour:2016rkg}, class Ia DHOST theories can be disformally transformed to Horndeski. It's proved in Appendix \ref{apdx:disformal} that such field redefinition is ill-defined in the example considered here.

\section{Discussion}\label{sec:dis}

Currently, GR is the well-tested effective theory of gravity.
Based on the higher-order derivative operators, which might
capture the physics of a UV-complete theory, we propose a
consistent EFT
\begin{equation}\label{GRcov}
\bal L=& {M_P^2\over 2}\,^{(4)}R+ P(\phi,X) -\left( B+{X^2
B^2\over 4M_P^2} \right) \phi^\mu \phi_{\mu \nu} \phi_\lambda
\phi^{\lambda \nu}  + B\phi^\mu \phi^\nu \phi_{\mu \nu} \Box \phi
\\ &+ {X B^2\over M_P^2}(\phi_\mu \phi^{\mu \nu } \phi_\nu)^2,
\eal
\end{equation}
for the spatially-flat fully stable nonsingular cosmologies. It
belongs to a subclass ($A=M_P^2/2$, $Q=0$) of the $c_T=1$ DHOST theory
\eqref{ct1 dhost cov}. It has been speculated that the
higher-order derivative operators ${\cal O}\left((\Box\phi)^2,
(\phi^{\mu\nu})^2\cdots\right)$ in the EFT \eqref{effective} might
play crucial roles in nonsingular cosmologies. Here, we clearly
showed what kind of ${\cal O}\left((\Box\phi)^2,
(\phi^{\mu\nu})^2\cdots\right)$ is required for the full stability
of nonsingular cosmologies.

We discussed how to evade the No-go Theorem with the EFT
\eqref{GRcov} (its ADM Langrangian \eqref{ct1 dhost adm1}). In
Refs.\cite{Cai:2016thi,Creminelli:2016zwa,Cai:2017dyi,Kolevatov:2017voe},
the operator ${\cal R}\delta g^{00}$ is used to expel the
No-go. However, in their implementation, besides higher-order derivative operators, the
corresponding covariant EFT also includes the derivative coupling
of $\phi$ to gravity $\sim X\,^{(4)}R$. Here, we found that the No-go can be evaded solely by introducing the higher-order
derivative operators ${\cal O}\left((\Box\phi)^2,
(\phi^{\mu\nu})^2\cdots\right)$ (the DHOST operators) in
\eqref{GRcov} without modifying GR. A concrete model of the
cosmological bounce have been presented in Sect.\ref{sec:model}. Generally, all the operators compatible with the symmetry of the problem are expected to be generated at quantum level. However, only a finite subset of all possible higher-order derivative operators is considered in the example studied. It would thus be interesting to study whether such model is protected against quantum corrections \cite{Pirtskhalava:2015nla,Santoni:2018rrx}. It might be also interesting to apply the EFT \eqref{GRcov} to
regulate the singularity of the BH,
e.g.\cite{Mironov:2018pjk,Franciolini:2018aad,Mironov:2018uou}.

Recently, the well-posedness of the initial value problem (IVP) has been promoted in non-perturbative cosmologies \cite{Ijjas:2018cdm}.
An issue worthy of exploring is whether the IVP
for \eqref{GRcov} is well-posed.

\begin{figure}
    \centering
    \includegraphics[width=4in]{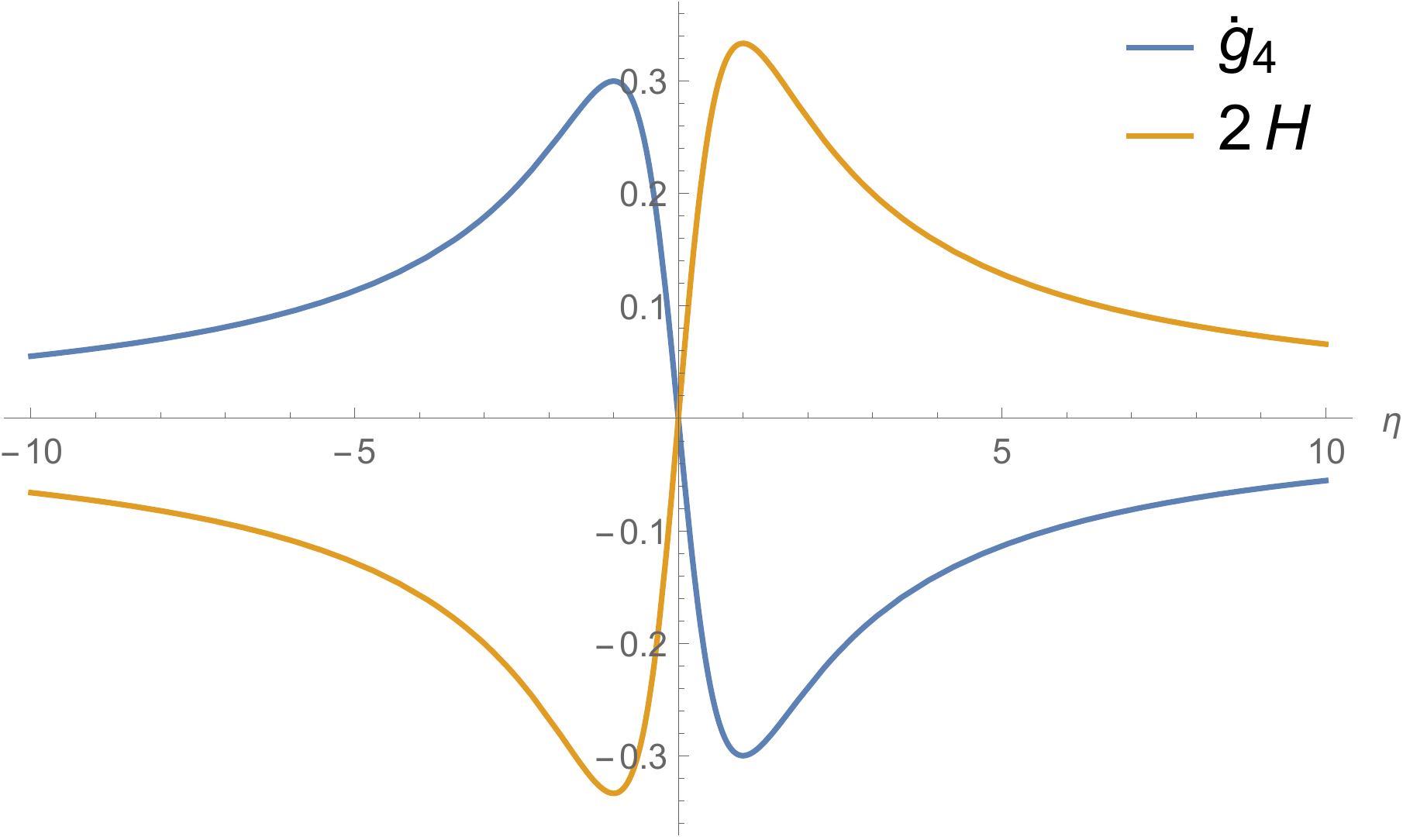}
    \caption{$\dot{g}_4$ is given by \eqref{g4}. We require ${\dot g}_4\sim H$ for simplicity. We have set $\mu=0.9$ and $M_P=10$ in the plot.}
    \label{H g4}
\end{figure}

%Now we turn to the asymptotic behavior of $g_1$ and $g_2$. Assuming that $c_1(\e)\to0$ exponentially as $|\e|\to\infty$, the asymptotic forms are obtained by setting $c_1=g_4=0$
%\begin{equation}\label{g1}
%\frac{g_1(\e)}{2N^2}\Big|_{\e\to+\infty}=-\frac{3M_p^2}{2N^2}\left(\h^2+\h'\right)\sim-\frac{1}{2}\frac{2M^2_p}{3}\frac{(\partial\phi)^2}{\phi^2}
%\end{equation}
%\begin{equation}\label{g2}
%\frac{g_2(\e)}{N^4}\Big|_{\e\to+\infty}=\frac{M^2_p}{4N^4}\left(3\h^2+\h'\right)\sim\frac{M^2_p}{4\e^2}\frac{(\partial\phi)^4}{\phi^4}
%\end{equation}
%\eqref{g1} is canonicalized by field redefinition $\pi=M_p\sqrt{2/3}\log\phi$ while the ghost condensate term \eqref{g2} vanishes in the future asymptotic\footnote{$g_2$ tends to zero much faster than $g_1$ because the leading time dependence of $H$ in the remote future is in the classically favored form $1/(3t)$. The extra $\e^{-2}$ suppression is because the next leading order term of $\h=\frac{\e^3}{3(1+e^4)}$ is proportional to $\e^{-5}$.}.

\textbf{Acknowledgments}

We thank Yong Cai for helpful discussions. This work is supported
by NSFC, Nos.11575188, 11690021.

\appendix
\section{The background equations}\label{apdx:EoM}
Varying \eqref{ct1 dhost adm} with respect to $N$ and $\h$,
respectively, we get
\begin{equation}\label{eom1}
\begin{aligned}
3B\left(\frac{ A B \ddot{N}}{N}+\frac{2 A
\dot{\h}}{N}\right)=&\dot{N}^2 \left(\frac{3 A B^2}{2 N^2}-\frac{3
A B B_N}{N}-\frac{3 A_N B^2}{2 N}\right)+\h^2 \left(-\frac{18 A
B}{N}-\frac{6 A}{N^2}+\frac{6 A_N}{N}\right)\\&+\dot{N}
\left(-\frac{9 A B^2 \h}{N}-\frac{6 A B B_\e}{N}-\frac{3 A_\e B^2}{N}\right)\\&+\h \left(-\frac{6 A B_\e}{N}-\frac{6
A_\e B}{N}-3 Q_N\right)-P_N N-P,
\end{aligned}
\end{equation}
\begin{equation}\label{eom2}
\begin{aligned}
6\left(\frac{ A B \ddot{N}}{N}+\frac{2 A
\dot{\h}}{N}\right)=&\dot{N}  \left(\frac{12 A\h}{N^2}-\frac{12
A_N\h}{N}-\frac{6 A B_\e}{N}-\frac{6 A_\e B}{N}+3
Q_N\right)\\&+\dot{N}^2 \left(\frac{9 A B^2}{2 N}+\frac{6 A
B}{N^2}-\frac{6 A B_N}{N}-\frac{6 A_N B}{N}\right)\\&-\frac{18 A
\h^2}{N}-\frac{12 A_\e \h}{N}-3 P N+3 Q_\e.
\end{aligned}
\end{equation}
%Equations \eqref{eom1} and \eqref{eom2} are clearly degenerate since the dynamical parts ($\ddot{N}$ and $\dot{\h}$) are the same while the RHS in both equations only contain $\dot{N}$, $N$ and $\h$.

\section{Solutions of $g_1$, $g_2$ and $g_3$}\label{apdx:g123}

\begin{equation}\label{g1}
\begin{aligned}
g_1=-\frac{1}{8 N}\big(&4 c_1 \h M_p^2 N^4 \dot{g}_4+2 c_1 M_p^2
g_4 N^4 \dot{g}_4 {\dot N}+c_1 M_p^2 N^5 \dot{g}_4^2+4 c_1 \h
M_p^2 g_4 N^3 \dot{N}+c_1 M_p^2 g_4^2 N^3 \dot{N}^2\\&+4 c_1 \h^2
M_p^2 N^3-12 \h M_p^2 N^2 \dot{g}_4-36 M_p^2 g_4 N^2 \dot{g}_4
\dot{N}+12 M_p^2 N \dot{g}_4 \dot{N}+6 M_p^2 N^3
\dot{g}_4^2\\&-108 \h^2 M_p^2 g_4 N^2-36 M_p^2 g_4 \dot{\h} N^2-72
\h M_p^2 g_4^2 N^2 \dot{N}+24 \h M_p^2 g_4 N \dot{N}\\&-24 M_p^2
g_4^2 N^2 \ddot{N}+12 M_p^2 g_4 N \ddot{N}+18 M_p^2 g_4^2 N
\dot{N}^2-12 M_p^2 g_4 \dot{N}^2+24 \h^2 M_p^2 N\\&+24 M_p^2
\dot{\h} N-24 \h M_p^2 \dot{N}\big),
\end{aligned}
\end{equation}
\begin{equation}\label{g2}
\begin{aligned}
g_2=\frac{1}{32}\big(&4 c_1 \h M_p^2 N^5 \dot{g}_4+2 c_1 M_p^2 g_4
N^5 \dot{g}_4 \dot{N}+c_1 M_p^2 N^6 \dot{g}_4^2+4 c_1 \h M_p^2 g_4
N^4 \dot{N}+c_1 M_p^2 g_4^2 N^4 \dot{N}^2\\&+4 c_1 \h^2 M_p^2
N^4+12 \h M_p^2 N^3 \dot{g}_4-12 M_p^2 g_4 N^3 \dot{g}_4 \dot{N}+4
M_p^2 N^2 \dot{g}_4 \dot{N}+6 M_p^2 N^4 \dot{g}_4^2\\&-36 \h^2
M_p^2 g_4 N^3-12 M_p^2 g_4 \dot{\h} N^3-36 \h M_p^2 g_4^2 N^3
\dot{N}+24 \h M_p^2 g_4 N^2 \dot{N}\\&-12 M_p^2 g_4^2 N^3
\ddot{N}+4 M_p^2 g_4 N^2 \ddot{N}+18 M_p^2 g_4^2 N^2 \dot{N}^2-4
M_p^2 g_4 N \dot{N}^2+24 \h^2 M_p^2 N^2\\&+8 M_p^2 \dot{\h} N^2-8
\h M_p^2 N \dot{N}\big),
\end{aligned}
\end{equation}
\begin{equation}\label{g3}
\begin{aligned}
g_3=-\frac{1}{32 N^3}\big(&-4 c_1 \h M_p^2 N^4 \dot{g}_4-2 c_1
M_p^2 g_4 N^4 \dot{g}_4 \dot{N}-c_1 M_p^2 N^5 \dot{g}_4^2-4 c_1 \h
M_p^2 g_4 N^3 \dot{N}\\&-c_1 M_p^2 g_4^2 N^3 \dot{N}^2-4 c_1 \h^2
M_p^2 N^3+36 \h M_p^2 N^2 \dot{g}_4+60 M_p^2 g_4 N^2 \dot{g}_4
\dot{N}\\&+12 M_p^2 N \dot{g}_4 \dot{N}-6 M_p^2 N^3
\dot{g}_4^2+180 \h^2 M_p^2 g_4 N^2+60 M_p^2 g_4 \dot{\h}
N^2\\&+108 \h M_p^2 g_4^2 N^2 \dot{N}-24 \h M_p^2 g_4 N \dot{N}+36
M_p^2 g_4^2 N^2 \ddot{N}+12 M_p^2 g_4 N \ddot{N}\\&-42 M_p^2 g_4^2
N \dot{N}^2-12 M_p^2 g_4 \dot{N}^2+72 \h^2 M_p^2 N+24 M_p^2
\dot{\h} N-24 \h M_p^2 \dot{N}\big).
\end{aligned}
\end{equation}
In Sect.\ref{sec:model}, since $N=1$, \eqref{g1}, \eqref{g2} and
\eqref{g3} will be simplified.

\section{Disformal transformations}\label{apdx:disformal}
In this appendix, we will show that the field redefinition relating the example in Sect.\ref{sec:model} to a Horndeski theory is ill-defined. According to Ref.\cite{Achour:2016rkg}, theory \eqref{ct1 dhost cov} can be disformally transformed to a Horndeski theory by the field redefinition $\tilde{g}_{\mu\nu}=\Omega(X,\phi)g_{\mu\nu}+\Gamma(X,\phi)\phi_\mu\phi_\nu$ where \[
\frac{\Omega_X}{\Omega}=\frac{4A_X+BX}{4A}, \qquad \Gamma_X=\frac{2A_X\Omega-2A\Omega_X}{AX}.
\]
A necessary condition for an invertible disformal transformation is \cite{Langlois:2017mxy}
\[\Omega-X\Omega_X-X^2\Gamma_X\ne0 \ . \]
For the specific example studied in Sect.\ref{sec:model}, $A=M_p^2/2, \ B=-M_p^2N^5g_4, \ X=-1/\bar{N}^2=-1$, thus $\Omega-X\Omega_X-X^2\Gamma_X=(1-g_4/2)\Omega$. According to \eqref{g4}, the disformal transformation is singular at the bounce point.

\end{document}